\documentclass[11pt]{article}
\usepackage{graphicx}
\usepackage{amssymb}
\usepackage{amscd}
\usepackage{mathrsfs}
\usepackage{longtable,lscape}
\usepackage{amsthm}
\usepackage{amsfonts}
\usepackage{amsmath}
\usepackage{bbm}
\usepackage{float}
\usepackage{url}
\usepackage{hyperref}
\usepackage{csquotes}
\usepackage{wrapfig}
\usepackage[round]{natbib}

\usepackage{caption}
\usepackage{lineno}
\usepackage[title]{appendix}
\begin{document}
\title{\bf 
The nature of time and motion in relativistic 
\\ operational reality
}
\author{Diederik Aerts and Massimiliano Sassoli de Bianchi\vspace{0.5 cm} \\ 
\normalsize\itshape
Center Leo Apostel for Interdisciplinary Studies, \\ \itshape 
Vrije Universiteit Brussel, 1050 Brussels, Belgium\vspace{0.5 cm} \\ 
\normalsize
E-Mails: 
\url{diraerts@vub.be}, \ \url{msassoli@vub.be}
}
\date{}
\maketitle
\begin{abstract} 
\noindent 
We argue that the construction of spacetime is personal, specific to each observer, and requires combining aspects of both discovery and creation. What is usually referred to as the block universe then emerges by noting that part of the future is contained in the present, but without the limitations that the four-dimensional block universe usually implies, of a reality in which change would be impossible. In our operational approach, reality remains dynamic, with free choice playing a central role in its conceptualization. We therefore claim that Einstein's relativity revolution has not been fully realized, since most physicists do not seem to be open to the idea that objects move not only in space, but also and especially in time, and more generally in spacetime, with their rest mass being a measure of their kinetic time energy. When relativistic motion is revisited as a genuine four-dimensional motion, it becomes possible to reinterpret the parameter $c$ associated with the coordinate speed of light, which becomes the magnitude of the four-velocity of all material entities. We also observe that the four-dimensional motion in Minkowski space can be better understood if placed in the broader perspective of quantum mechanics, if non-locality is interpreted as non-spatiality, thus indicating the existence of an underlying non-spatial reality, the nature of which could be conceptual, consistent with the conceptuality interpretation of quantum mechanics. This hypothesis is reinforced by noting 
that when observers, or experiencers, as they will be referred to in this article, are described by acknowledging their cognitive nature, of entities moving in a semantic space, Minkowski metric emerges in a natural way.
\end{abstract} 
\\

\section{Introduction}
\label{introduction}

Reconciling the reality of one's present experience with that of the four-dimensional continuum of special relativity constitutes one of conceptual difficulties posed by Einstein's celebrated theory \citep{Einstein1905,einstein1920}. Indeed, physical entities are often imagined as moving along their worldlines, but then the following question arises: What truly exists, the entities in motion on their worldlines, or the worldlines themselves, or, does the question about ‘what truly exists’ need still another answer? Also, do all entities only exist in the present moment, according to the view of \emph{presentism}, or it is the view of \emph{eternalism} which is more correct, when it says that entities jointly exist in all their temporal dates, so not only in their present but also, jointly, in their past and future \citep{presentism}? 

Without a doubt, relativity fully brings into play this dichotomy between the opposing views of presentism and eternalism, confronting us with the question of knowing what truly exists, what \emph{change} really is, whether it really is such, or just an illusion. These questions, however, are not only the proper of relativity, they were, and still are, at the core of the research on the foundations of quantum mechanics, which can be said to have begun in the 1970s, following the critical reflection contained in the celebrated EPR article \citep{EPR1935}, which anticipated Bell's work \citep{bell1964} and his inequalities that, unexpectedly, allowed to experimentally test the reality of entanglement and non-locality \citep{freedmanclauser1972,weishetal1998}. 

More precisely, similar to what Einstein did in relativity with the measurement of distances and durations, in quantum mechanics it was also possible to operationally define \emph{what exists} from an analysis of the different measurement procedures and their relation to the notion of \emph{prediction}, viewing \emph{actuality} as a special \emph{state of prediction}, corresponding to the situation where the outcome of the experimental test of a property is 100\% certain, whereas \emph{potentiality} is characterized by a weaker \emph{probabilistic prediction}, i.e., a situation where the outcome in question has a probability strictly lower than 100\%, which cannot be made equal to 100\% even in principle \citep{Aerts1982,Aerts1983}.  This means that potentiality would have its proper place in reality, and how it is partly detected, or observed, 
and, in considering quantum theory, also partly constructed, or even created, an aspect of the situation that we will specify more fully later in this article. 

It is important to note that what exists does not necessarily allow itself to be circumscribed in purely operational terms. However, what exists in an operational sense must also exist in a broader metaphysical sense. In the way the theory of relativity was presented and derived by Albert Einstein himself \citep{einstein1920}, the notion of an observer plays an important role. However, even when using a clock and ruler to measure intervals of time and space, an observer acts and interacts with the reality he or she observes. Thus, he or she does much more than simply passively observe. Of course, these specific actions of measuring spatiotemporal intervals remain very close to the idea of passive observation, so the notion of observer still seems appropriate. But as will emerge more clearly in our analysis of relativity, inspired by our work on the foundations of quantum mechanics, in Einstein's observer there is a fundamental active aspect previously unnoticed, which is why we shall henceforth call an observer, more appropriately, an \emph{experiencer}.

Indeed, to operationally define what exists, we have to refer to the notion of \emph{experience}, which in turn depends on the \emph{personal power} of an experiencer, what he or she is in principle able to interact with, e.g., through his or her body and instruments. And since an experiencer's power to ``touch'' the real, both in width and depth, grows proportionally to his or her 
knowledge, the corresponding definition of operational reality will also grow accordingly \citep{Aerts1996}. 

Note however that our reality, in each moment, is not just the content of our experience in that moment, as if this would be true it would be extremely limited. It is, instead, the collection of all our \emph{possible} experiences in that moment, those we could have lived should we have made different decisions in our past. This shows the importance of \emph{free choice} in our reality construction, as well as the fact that an operational reality is a \emph{personal reality}, which is constructed individually by each observer. It is then natural to ask: Can we coherently integrate all the personal present realities associated with the different experiencers into a global present reality construction? 

Such a global construction out of local personal realities was what could be accomplished within the Newtonian worldview, using the existence of a single time flow, shared equally by all experiencers. In other words, an absolute Newtonian time, advancing inexorably in the same way for each of them. But special relativity tells us that the present is personal, that there is not one time, but multiple personal times, and this leads to surprises in our operational construction of reality, which is what we aim to explain and illustrate in this article, which is organized as follows. 

In Section~\ref{The future in the present}, we identify `what exists' in an operational sense in relativity and show that, as a consequence of the effect of time dilation, the future is literally also in the present, hence each observer, hence experiencer, is associated with a personal four-dimensional block universe. However, change remains natural and at the core of reality. Indeed, as our analysis will make clear, it is the Newtonian reflex of wanting to fuse these personal block universes into a single global construction that causes the problem with change, in the way we usually reflect on the notion of block universe, hence the problem of eternalism.

In Section~\ref{The proper speed of light}, as a further deepening and fine tuning of this view, we revisit the notion of coordinate velocity, emphasizing that the notion of \emph{proper velocity}, or \emph{celerity}, is more adequate to describe the spatial motion of physical entities, making the invariance of the speed of light much more intuitive. Continuing our analysis, in Section~\ref{Moving with a four-velocity}, we observe that the notion of proper velocity is part of a more general notion, that of \emph{four-velocity}, which allows us to reinterpret the structural parameter $c$ appearing in the Lorentz transformation as the absolute speed of all material entities. 

In Section~\ref{Multiplicity of times}, we observe how the existence of a multiplicity of proper times implies that the block universe strategy of conferring the worldlines and worldtubes an intersubjective reality fails in the same way that simultaneity fails in relativity. In Section~\ref{Temporal energy}, we show that the movement along the time direction can be associated with a kinetic-like energy, which is nothing other than the mass energy of a physical entity. In Section~\ref{Quantum and conceptuality}, we briefly introduce the perspective of the \emph{conceptuality interpretation of quantum mechanics}, describing quantum non-locality as non-spatiality and conceptuality, and in Section~\ref{Explaining Minkowski}, we use the conceptuality input to derive the Minkowski metric in a very natural way. Finally, in Section~\ref{Conclusion}, we recapitulate our findings, offering some final remarks.

\section{The future in the present}
\label{The future in the present}

The starting point in the definition of the \emph{present personal reality} of an observer, and we mean here the notion observer as used in Einstein's version of relativity \citep{einstein1920}, 
is the notion of \emph{experience} \citet{Aerts1996,Aerts1999}. The general situation is that such an observer only has one experience at a time, i.e., there is only one \emph{present personal experience}, and of course, when we say `present', we are referring here to the \emph{proper time} of the observer. 

There are two fundamental aspects in an experience: a \emph{creation aspect}, and a \emph{discovery aspect}. The former is that aspect of an experience that is acted upon by the observer, 
whereas the latter is that aspect of an experience that lends itself to such action-creation, being present independently of such action and which, therefore, can be discovered while performing it. Let us call this second aspect a \emph{happening}.  We could have used the notion of \emph{event} instead of happening, to indicate this second aspect, but the general consensus in identifying an event with a point in spacetime, that is, an element of Minkowski space, does not make this notion general enough. Our intention is to introduce a framework in which quantum mechanics also finds a place, so that we can construct a theory that fully reconciles relativity theory and quantum mechanics. We have just mentioned that in our approach quantum non-locality is interpreted as non-spatiality, so, limiting `what exists' to `spacetime events' is not an option, hence the use of the more general notion of happening instead of event.

In our language, creations are usually expressed by verbs and happenings by substantives. The crucial point is that although an observer may have only one present experience, there are many experiences that he/she could have had in replacement of his/her present experience, if only he/she had made different choices in his/her past. This because many other happenings are also available, in that same moment, to be part of his/her present experience, and their collection is by definition the \emph{present personal reality} of the observer. It is this freedom of choice, that is, the existence of the possibility for an observer to experience something different by making a different choice in the past, that is crucial to the operational construction of `what exists’. This hypothesis of freedom of choice is not made explicit in Einstein's version of relativity \citep{einstein1920}, although it is deeply linked to the scientific project itself and its operational foundation. It is precisely in order to emphasize the importance of this possibility of free choice for an observer that we have decided to introduce, as already announced, the new notion of an experiencer, with the usual relativistic observer who can be seen as a simplified and idealized version of this more general experiencer, whose non-quantum reality, at a given moment in time, is considered only a spatial reality.

Note also that the notion of event, although within standard approaches to relativity it is considered very general, is rather limited, indicating a passive worldview, in which events simply happen and are possibly observed. Our approach, inspired by the foundations of quantum theory, adopts a non-passive worldview, where an observation is considered not only an act of discovery, but also of creation, as evidenced in the quantum formalism by the process of the wavefunction collapse following a measurement. This is why we consider an observer also 
as an experiencer. Hence, in what follows we will adopt an idealization like that adopted in standard approaches to relativity, considering only that subclass of happenings reducible to spatiotemporal elements. However, we will suppose that an experiencer has several happenings at his or her disposal, and thus there is freedom of choice for an experiencer. 

So, one can associate to each relativistic experiencer 
a \emph{present personal space}, which is a special subset of his/her \emph{present personal reality}, containing all the happenings that are available to the experiencer 
at that moment. But even when we limit our analysis to happenings that `happen in space', the present personal reality of a given experiencer  is more complex than we are usually used to consider. Indeed, although the personal space of a given experiencer, defined by his/her proper reference frame, is a \emph{space of simultaneity}, such simultaneity is only relative to the 
experiencer's clock, his/her personal spatial reality being also populated by entities existing in multiple temporal versions, and in that sense, it is truly a four-dimensional realm, i.e., a spacetime.

To explain this, consider two clocks, let us call them clock-A and clock-B. Let us assume that an experiencer  
named Alice is sitting in her office and that clock-A is in her pocket, whereas clock-B is in her desk drawer. They both mark the same time, let's say 13:00:00, which is the present (proper) moment of Alice. In other words, clock-A and clock-B, both marking time 13:00:00, are jointly part of Alice's present personal reality at time 13:00:00. As we explained, this is so because both clocks are available to become part of Alice's experience. For instance, at time 12:59:55, Alice could have either decided to take clock-A out of her pocket and look at it (assuming the operation takes 5 seconds), discovering in this way that it marks 13:00:00, or she could have decided to take clock-B from the drawer (assuming again that the operation takes 5 seconds) and look at its face to also discover that it marks exactly 13:00:00. 

It is also true, however, that an hour before, at time $t_{\rm p}^1=\,$11:59:55, Alice could have traveled back and forth along a given spatial direction. Let us say that she could have done so at \emph{proper space velocity} $v_{\rm p}$ (see the next section for its definition) and let us assume for simplicity that she can elastically revert her path after exactly half an hour. Note that we have described an action of Alice that is as simple as possible, as is customary in relativistic texts, but of course it may be more complex in terms of spatial trajectory. What is important is that she moves away from her office and then after a certain time returns, and in modeling her action the accelerations that she necessarily experiences are neglected. 

\begin{figure}[!ht]
\centering
\includegraphics[scale =0.3]{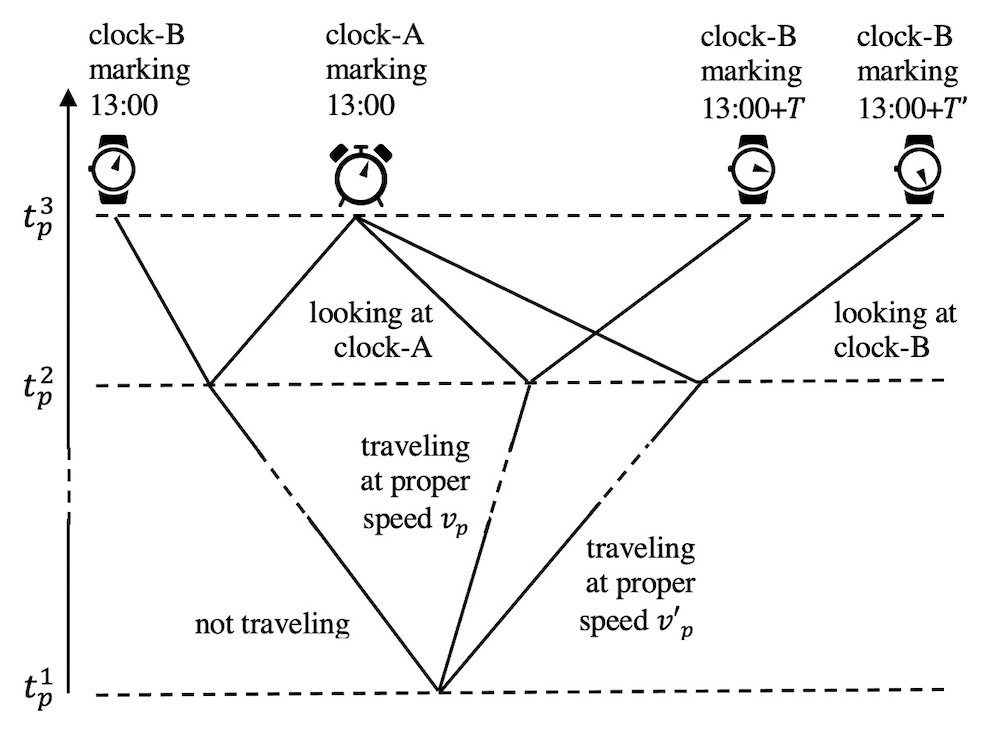}
\caption{At time $t_{\rm p}^1=\,$11:59:55, Alice can either remain in her office, or travel back and forth, with different proper space velocities, for example $v_{\rm p}$ or $v'_{\rm p}$, being back at time $t_{\rm p}^2=\,$12:59:55. At this point, she can decide to look at either clock-A or clock-B. The former will show the time $t_{\rm p}^3=\,$13:00:00, which is Alice's proper time, whereas the latter will show a time which depends on Alice's past decisions. If she decided not to travel, it marks 13:00:00; if she decided to travel at proper space velocity $v_{\rm p}$, it marks 13:00:00$\,+T$, with $T=(\gamma_{\rm p}-1)(t_{\rm p}^2-t_{\rm p}^1)$; if she decided to travel at proper space velocity $v'_{\rm p}$, it marks 13:00:00$\,+T'$, with $T'=(\gamma'_{\rm p}-1)(t_{\rm p}^2-t_{\rm p}^1)$, and so on. This means that countless versions of clock-B, of different ages, jointly exist in Alice's present personal reality, as a consequence of the relativistic effect of time dilation. (Note that the above is not a spacetime diagram, but a bifurcation diagram, where Alice's proper time plays the role of the bifurcation parameter). 
\label{figure1}}
\end{figure}

Now, if Alice would have done so, when back at her office at personal time $t_{\rm p}^2=\,$12:59:55, she could also have decided to either look at clock-A in her pocket, discovering that it marks $t_{\rm p}^3=\,$13:00:00, or to take clock-B from the drawer, and by looking at it she would have discovered that it doesn't mark 13:00:00, like her clock in the pocket, but 13:00:00$\,+\,T$, where 
\begin{equation}
T=(\gamma_{\rm p}-1)(t_{\rm p}^2-t_{\rm p}^1)\quad\quad \gamma_{\rm p}=\sqrt{1+{v_{\rm p}^2\over c^2}}
\label{T}
\end{equation} 
Since $\gamma_{\rm p}$ can take values from $1$ to infinity, when $v_{\rm p}$ varies from $0$ to infinity, we obtain that not only clock-B marking 13:00:00 is part of Alice's personal reality at time 13:00:00, but also clock-B marking any time from 13:00:00 to infinity, or to be more precise, any time from 13:00:00 up to the time that corresponds to the end of the existence of clock-B as a physical entity with limited life span; see Figure~\ref{figure1}.

In other words, if it is true that, for how we build our physical reality, the personal space of Alice at a given personal moment $t_{\rm p}$, is a three-dimensional manifold formed by all the possible spatiotemporal happenings (events) associated with the same temporal order parameter $t_{\rm p}$,  when we also consider the personal times of the entities populating it, we obtain a genuine four-dimensional description. This because an entity like a clock, or any other material entity that can ideally be associated with a spatial coordinate, are happenings that are available to be experienced in an infinite number of temporal versions, compatibly with their lifespans and with the fact that nothing has intervened, at a given moment of their existences, to destroy them. So, when operationally defining what is real for a given experiencer, in a given moment of his/her proper-personal time, by means of the collection of all the happenings that can be fused with his/her creations at that time, within the limits of his/her personal power, one comes to the deep insight that even things that one would normally situate in one's personal future are existing in one's personal present. 

Before continuing our analysis, a few remarks are in order. We can observe that the Minkowski four-dimensional spacetime structure only contributes to the possible experiences of an  
experiencer on the happening side, i.e., it only describes the spatial happenings that are available to be fused with one of the experiencer's  creations, to live a given experience, in a given moment (like the experience of looking at a watch). All these happenings, which jointly exist in that moment, in countless temporal versions, give rise to a genuine four-dimensional spacetime structure.

The statement  that clock-A marking 13:00:00, and that same clock marking 14:00:00, can be both part of Alice's personal reality at time 13:00:00, 
may lead to believe that the future would be in close proximity to the present, since these two versions of clock-A are located, spatially speaking, in the same place (the drawer). But this would be an incorrect conclusion, as in fact clock-A marking 14:00:00 is very far away from Alice. Indeed, to be able to have an experience with it, at personal time 13:00:00, she had to travel at very high proper space velocity. 

Note also that if it is true that happenings of different ages jointly exist in an experiencer's reality, only one temporal version of the experiencer exists, in a given moment, as is clear that no action can be performed by the latter, in his/her past, that would allow him/her to observe himself/herself with a different personal time than his/her actual personal time. This is also the reason why we can consider the above mentioned four-dimensional spacetime structure to exist only from the perspective of the experiencer, in the sense that the experiencer himself/herself is not contained into it. In other words, his/her four-dimensional \emph{personal block universe} exists at a given moment of his/her proper time as the collection of what is real to him/her, in terms of events, in that moment, excluding himself/herself from that collection.

This is how we believe the notion of a personal block universe should be precisely understood, as something personal to a given experiencer, in the same way proper times and proper velocities are. It is the collection of all the existing worldlines (or worldtubes, when considering objects that are not point-like) whose lengths depend on the lifespan of the entities that generate them. And the same applies to experiencers other than the one under consideration, who again are not part of their own personal block universes. But then, what is it that truly exists, the worldtubes or the entities moving along them? Our approach provides a rather nuanced answer to this question, being clear that it depends on the perspective adopted, since when we talk about the reality of an experiencer, it does not contain the entity that is the subject of the experiences that underpin its construction, while it includes  the worldtubes corresponding to the other experiencers.

Note that the existence of personal block universes does not imply that change would be impossible. Coming back to Alice, her reality is dynamic, the four-dimensionality of the entities (different from her body) being a consequence of the fact that, through her free choice, she can select her own possible experiences, enabling her, via the time dilation effect, to possibly travel into the future of other entities, and experience them there. From a quantum perspective, if Alice's reading of clock-B is viewed as a \emph{measurement}, then her possible round trips correspond to different possible \emph{preparations} of the state of the measured system, i.e., clock-B. And the fact that time and reality become personal in relativity, this can also be viewed as that typical quantum-mechanical feature called \emph{contextuality}, although we have here a specific relativistic type of spatiotemporal contextuality. 

We also observe that the existence of this collection of personal block universes does not give rise to the existence of a single block universe without experiencers, the situation with respect to the question `What exists when there are no experiencers?' being more complex than that. We will return to this issue later in our article. But let us already say that it is not our aim to develop an idealist interpretation of the theory of relativity, in the sense that `observation' would be necessary for `existence'. Our operational approach is mainly aimed at revealing the nature of reality, that is, of what exists, considering how we access knowledge and structure it. For example, we take for granted that nature and its relativistic properties exist even when no one 
experiences them, and already existed when there were no human beings to experience them. When in Section~\ref{Quantum and conceptuality} we briefly introduce our conceptuality interpretation of quantum theory, we will still be able to express a more nuanced view on this and other related issues.

\section{The proper speed of light}
\label{The proper speed of light}

When we consider spacetime and the block universe in personal terms, i.e., as a personal construction proper to each reality's experiencer, we are in fact simply extending those personal notions that are already present in the relativity textbooks. Think of \emph{proper time} and \emph{proper length}, which are clearly personal to a given experiencer. However, remnants of a pre-relativistic (Newtonian) thinking are still present in those relativity textbooks, with the risk of obscuring what the formalism seeks to reveal to us, if we only choose to take it seriously. 

We mentioned already one of them, namely wanting to think of the different personal block universes as if they were one single global block universe. But how are these different personal block universes related then? It is the coordinate velocities that play this role of relating the different experiencers, contributing to the structure of global reality, together with the free choices that experiencers can make at every personal instant, according to their personal power of fusing one of their creations with a selected happening, to bring an experience to life. 

It is worth mentioning here the hypothesis of \emph{superdeterminism} \citep{Brans1988,Sabine2020}, according to which freedom of choice would not exist. If that hypothesis is true, our operational construction would not work. Indeed, it would mean that only one experience is possible at any instant. But if free choice does exist, the structure of global reality necessarily contains countless bifurcations towards the future, hence is very different from a unique block universe. 

That said, to shed some light on another major Newtonian bias, consider the notion of \emph{proper velocity}, $v_{\rm p}$, also named \emph{celerity}, the magnitude of which is not limited, as it can go from zero to infinity, contrary to \emph{coordinate velocity}, $v$, whose magnitude can only go from zero to $c=299\, 792\, 458\, {\rm m/s}$. Proper velocity is rarely used for interpretative purposes, or in formulae, like for example in writing Lorentz transformations. But when we reason in terms of proper velocities, we find that light possesses an infinite proper speed, which allows one to understand the counterintuitive fact that light's coordinate speed is measured always equal to $c$, independently of the velocity of the source or of the experiencer  \citep{Einstein1905}. 

To see this in some detail, let us consider two spatiotemporal frames of reference, $\Sigma(t,x)$ and $\Sigma'(t',x')$, with the second frame moving at coordinate velocity $V$ with respect to the first. Not to complicate our discussion unnecessarily, we will only consider one spatial dimension, therefore, time, position and velocity will all be scalar quantities in our discussion. To further simplify, let us also assume that the origins of the two frames, $x = x' = 0$, coincide at times $t = t' = 0$. The transformations to go from $\Sigma(t,x)$ to $\Sigma'(t',x')$, called \emph{Lorentz boosts} are then given by:  
\begin{equation}
t'=\Gamma(t-\frac{V}{c^2}x)\quad\quad x'=\Gamma(-Vt+x)\quad\quad \Gamma={1\over \sqrt{1-{V^2\over c^2}}}
\label{Boost1}
\end{equation} 
where $\Gamma$ is the \emph{Lorentz factor}. 
Similarly, the inverse transformations, to go from $\Sigma'(t',x')$ to $\Sigma(t,x)$, are obtained by considering the change $V\to -V$ in the above formulae, which gives: 
\begin{equation}
t=\Gamma(t'+\frac{V}{c^2}x')\quad\quad x=\Gamma(Vt'+x')
\label{Boost2}
\end{equation} 

Suppose now that a body moves with coordinate velocity $v$ with respect to $\Sigma(t,x)$. To know the coordinate velocity it moves with respect to $\Sigma'(t',x')$, let us call it $v'$, we have to derive the \emph{relativistic composition law for coordinate velocities}. Using the above transformations, we obtain: 
\begin{equation}
v'= {dx'\over dt'}= {\Gamma(-Vdt+dx)\over \Gamma(dt-\frac{V}{c^2}dx)}={v-V\over 1 - \frac{Vv}{c^2}}=c\,{v - V\over c-{Vv\over c}}
\label{Variation}
\end{equation} 
where $v= {dx\over dt}$. When the magnitude of the velocities involved are small compared to $c$, the denominator in (\ref{Variation}) is approximately equal to $1$, and the formula reduces to the additive \emph{Galilean composition law for coordinate velocities}: 
$v'\approx v-V$. In other words, the theory of relativity tells us that the linear Galilean law is only an approximation of a more general non-linear law. 

One of the surprising aspects of the relativistic law is that in the limit $v\to c$, we obtain $v'\to c$. This means that everything happens as if relative motions did not matter in the limit where the speed of the body approaches $c$, in the sense that in both frames, $\Sigma(t,x)$ and $\Sigma'(t',x')$, it is observed to be exactly the same. The above  formulae tell us why this must be the case, but how can we understand this phenomenon on a more fundamental level? 

For this, it is important to remember that the revolution of the passage from Galilean relativity to Einsteinian relativity brings with it the passage from a single time, valid for every inertial experiencer, to a multiplicity of different times, which can be associated with the different inertial experiencers, and more generally with the different physical entities. When we calculate a velocity, a new problem therefore arises, which can be expressed with the following question: With respect to which temporal variation should one calculate the variation of the position of a physical entity?

Let us consider the simple example of a car. Every modern motor vehicle is equipped with a so-called \emph{speedometer}, an on-board instrument that allows to measure the distance traveled per unit time. But which time are we talking about here? Obviously, that measured by the speedometer's clock, which is part of the car and travels with it. To describe this situation, we now also consider the reference frame $\Sigma_{\rm p}(t_{\rm p},x_{\rm p})$ associated with the center of mass of the car moving at coordinate speed $v$ with respect to the reference frame $\Sigma(t,x)$, associated with the road. Since by definition the car's center of mass is at rest at the origin of $\Sigma_{\rm p}(t_{\rm p},x_{\rm p})$, Lorentz transformations 
(\ref{Boost2}) reduce to:
\begin{equation}
t=\gamma t_{\rm p}\quad\quad x=\gamma vt_{\rm p} \quad\quad \gamma=\frac{1}{\sqrt{1-\frac{v^2}{c^2}}}
\label{Boost-proper}
\end{equation}
where $t_{\rm p}$ is called {\it proper time} in relativity, and we have assumed that $x=x_{\rm p}=0$ at $t=t_{\rm p}=0$. The first of the above two identities is the \emph{time dilation} relation, whereas by deriving the second identity with respect to $t_{\rm p}$, we find that the {\it proper velocity} $v_{\rm p}=\frac{dx}{dt_{\rm p}}$, i.e., the velocity measured by the speedometer, also called {\it celerity}, is given by
\begin{equation}
v_{\rm p}=v \gamma =\frac{v}{\sqrt{1-\frac{v^2}{c^2}}}
\label{proper-velocty2}
\end{equation}
Clearly, when $|v|$ is small compared to $c$, we have the approximation $v_{\rm p}\approx v$. We also observe that when $|v|\to c$, $\gamma\to\infty$ and $|v_{\rm p}|\to\infty$. In other words, if we measure the speed of light with a ``speedometer protocol,'' i.e., if we measure the \emph{proper spatial velocity of light}, it no longer has a finite value, but an infinite one. 

We can also consider the \emph{composition law for proper velocities}. For this, we also introduce the proper velocity $V_{\rm p}=\Gamma V$ of the reference frame $\Sigma'(t',x')$ with respect to $\Sigma(t,x)$, and the proper velocity $v'_{\rm p}=\gamma' v'$ of the entity under observation relative to the reference frame $\Sigma'(t',x')$, with $v'_{\rm p}=\frac{dx'}{dt_{\rm p}}$. Replacing $\frac{v}{c}=\tanh{\vartheta}$ in (\ref{proper-velocty2}), we find that ${v_{\rm p}\over c}=\sinh{\vartheta}$, hence $\vartheta=\sinh^{-1}{\frac{v_{\rm p}}{c}}$, and similarly $\vartheta'=\sinh^{-1}{\frac{v'_{\rm p}}{c}}$, $\Theta=\sinh^{-1}{\frac{V_{\rm p}}{c}}$. Considering that $\vartheta'=\vartheta - \Theta$, we deduce the composition law: 
\begin{equation}
v_{\rm p}^\prime=c\sinh{\left(\sinh^{-1}{\frac{v_{\rm p}}{c}}-\sinh^{-1}{\frac{V_{\rm p}}{c}}\right)}
\label{proper-composition2}
\end{equation}
We can further transform this expression using the identities $\sinh{(\alpha-\beta)}=\sinh{\alpha}\cosh{\beta}-\cosh{\alpha\sinh{\beta}}$ and $\cosh\left(\sinh^{-1}{a}\right)=\sqrt{1+a^2}$, which becomes: 
\begin{equation}
v_{\rm p}^\prime=v_{\rm p}\sqrt{1+\frac{V_{\rm p}^2}{c^2}}-V_{\rm p}\sqrt{1+\frac{v_{\rm p}^2}{c^2}}
\label{proper-composition4}
\end{equation}
If the proper velocities involved are small compared to $c$, we 
recover the Galilean composition law $v_{\rm p}^\prime\approx v_{\rm p}-V_{\rm p}$, and if instead we let $|v_{\rm p}|\to\infty$, we have the asymptotic form:
\begin{equation}
v_{\rm p}^\prime=v_{\rm p}\left(\sqrt{1+\frac{V_{\rm p}^2}{c^2}}-\text{sign}(v_{\rm p})\frac{V_{\rm p}}{c}\right)+O(v_{\rm p}^{-1})
\label{proper-composition5}
\end{equation}

So, when using the notion of proper velocity, the counter-intuitiveness of the invariance of the speed of light disappears. Indeed, for photons and other zero-rest mass entities, the magnitude of their proper velocities is infinite, compatibly with (\ref{proper-composition5}), since multiplying a (positive or negative) infinity by a positive constant has no effect:
\begin{equation}
\pm\infty=\pm\infty\left(\sqrt{1+\frac{V_{\rm p}^2}{c^2}}-\text{sign}(v_{\rm p})\frac{V_{\rm p}}{c}\right)
\label{proper-composition-infinite}
\end{equation}
Now, if $v_{\rm p}$ is the notion one should use to characterize the spatial movement of an entity, consistent with our previous operational construction of reality, where time is something strictly personal, it follows that the historical coordinate velocity $v$ would provide a misleading representation of the movement of a physical entity, but considering that there is no difference between $v_{\rm p}$ and $v$ in the non-relativistic regime, we had no way of becoming aware of the problem before the advent of relativity. More precisely, the coordinate velocity $v$ would contain an undue relativistic contraction, as is clear that, reversing (\ref{proper-velocty2}), we have:
\begin{equation}
v={1\over \sqrt{1+{v_{\rm p}^2\over c^2}}} \,v_{\rm p}
\label{contraction}
\end{equation}

We can then say that because of the contraction (\ref{contraction}), which we were not aware we were performing in pre-relativistic times, a quantity whose range should naturally go from $0$ to $\infty$ (celerity $v_{\rm p}$), was contracted into a quantity with a bounded interval going from $0$ to $c$ (coordinate velocity $v$). In other words, the coordinate speed of light is always equal to $c$, which is the limit value of $|v|$, because it would not be its true spatial speed, but an infinite contraction of it, perfectly calibrated to always obtain the same finite value $c$, equal to $299\, 792\, 458\, {\rm m/s}$.

\section{Moving with a four-velocity}
\label{Moving with a four-velocity}

As a natural continuation and deepening of our analysis, we can observe that the proper velocity of a body also corresponds to the spatial component of a more general \emph{four-velocity}, a notion introduced in every textbook of relativity. Surprisingly, the magnitude of the four-velocity, in any reference system, is always exactly equal to $c$, and although this is a known result, it is usually regarded as a mere mathematical property with no physical meaning. But considering that Lorentz transformations can be derived without using Einstein's second postulate \citep{Ignatowsky1910,FrankRothe1911,Leblond1976}, the structural parameter $c$ appearing in them can be given a more general interpretation, as the \emph{absolute speed of all material entities}, whose motions occur in the entire spacetime, since they are always characterized by a nonzero time component.

Let us see this in some detail, always limiting our discussion to a single spatial dimension, hence the four-velocity will be here a two-velocity, with a time component and a single spatial component, but for clarity we will keep calling it four-velocity. More precisely, the four-velocity of an entity, relative to a reference frame $\Sigma(t,x)$, is given by the (here two-dimensional) vector \citep{Minkowski1908}:
\begin{equation}
u_{\rm p}= \left[\begin{matrix}v^0_{\rm p}\\ v_{\rm p}\\\end{matrix}\right]= \left[\begin{matrix}c\frac{dt}{dt_{\rm p}}\\ \phantom{c}\frac{dx}{dt_{\rm p}}\\\end{matrix}\right] = \left[\begin{matrix}c\gamma\\ v\gamma\\\end{matrix}\right] 
\label{2-velocity}
\end{equation}
To calculate its magnitude $\|u_{\rm p}\|$, one has to use the Minkowski metric, which has a minus sign for the spatial variables and a plus sign for the time variable. This gives:
\begin{equation}
\|u_{\rm p}\|=\sqrt{(v^0_{\rm p})^2-(v_{\rm p})^2}=\sqrt{\left(c\gamma\right)^2-\left(v\gamma\right)^2}=\gamma\sqrt{c^2-v^2}=c
\label{2-velocity-length}
\end{equation}
As we said, physicists do not pay much attention to this remarkable result. However, if we dare to take it seriously, it tells us something fundamental, namely that every physical entity always moves, relative to its personal block universe, hence on its worldline, at the same (proper) speed, which has the same value as the coordinate speed of light $c$. They do so ``always" in the sense that the magnitude of the four-velocity does not depend on the choice of the reference frame, hence, it is truly an intrinsic property of the moving entity. 

More specifically, the spatiotemporal (proper) velocity $u_{\rm p}$ has two components: a temporal component, $v_{\rm p}^0$, corresponding to the \emph{proper time velocity} with which the entity moves along the temporal axis, and a spatial component, $v_{\rm p}$, corresponding to the \emph{proper space velocity} with which it moves along the spatial axis (or spatial axes, if there is more than one spatial dimension). The proper time velocity $v_{\rm p}^0 = \gamma c$ is always positive, and we can interpret this as an indication that we can never go back in time, whereas the proper space velocity $v_{\rm p} = \gamma v$ has the sign of $v$, which can be both positive and negative, depending on the direction of motion along the $x$-axis. 

Furthermore, as we have already emphasized, $|v_{\rm p}|$ can take values ranging from $0$ to infinity, when $|v|$ goes from $0$ to $c$. The proper time velocity can also reach infinity, when $v$ tends to $c$, but its minimum value, which is obtained when $v$ tends to $0$, is $c$. If $v = 0$, this means that the entity in question is spatially at rest with respect to the reference frame $\Sigma(t,x)$. But even when spatially at rest, temporally it will never be at rest, since its proper time velocity is then equal to $c$. 

If we compare the magnitudes of the temporal and spatial components of the four-velocity $u_{\rm p}$, we immediately see from (\ref{2-velocity}) that in the non-relativistic regime, ${|v|\over c}\to 0$, the movement along the time-direction, $v_{\rm p}^0\to c$, dominates, as is clear that $v_{\rm p}^0-|v_{\rm p}|=v_{\rm p}^0(1-{|v|\over c})$. On the other hand, in the ultrarelativistic regime, ${|v|\over c}\to 1$, the magnitudes of the temporal and spatial movements become comparable, $v_{\rm p}^0-|v_{\rm p}|\to 0$, with both speeds tending to infinity, since $\gamma\to\infty$ when $|v|\to c$. 

The equality $|v|= c$ can only hold for entities of zero-rest mass, like photons. Hence, unlike material entities with non-zero rest mass, although they also move at spatiotemporal speed $c$, photons also possess infinite proper speeds along the temporal and spatial axes. But these two infinite values are such that they always compensate with precision when calculating the magnitude of the full spatiotemporal velocity (\ref{2-velocity-length}), thanks to the Minkowskian metric, to always yield the finite value $c$.

\section{Multiplicity of times}
\label{Multiplicity of times}

\begin{figure}[!ht]
\centering
\includegraphics[scale =0.2]{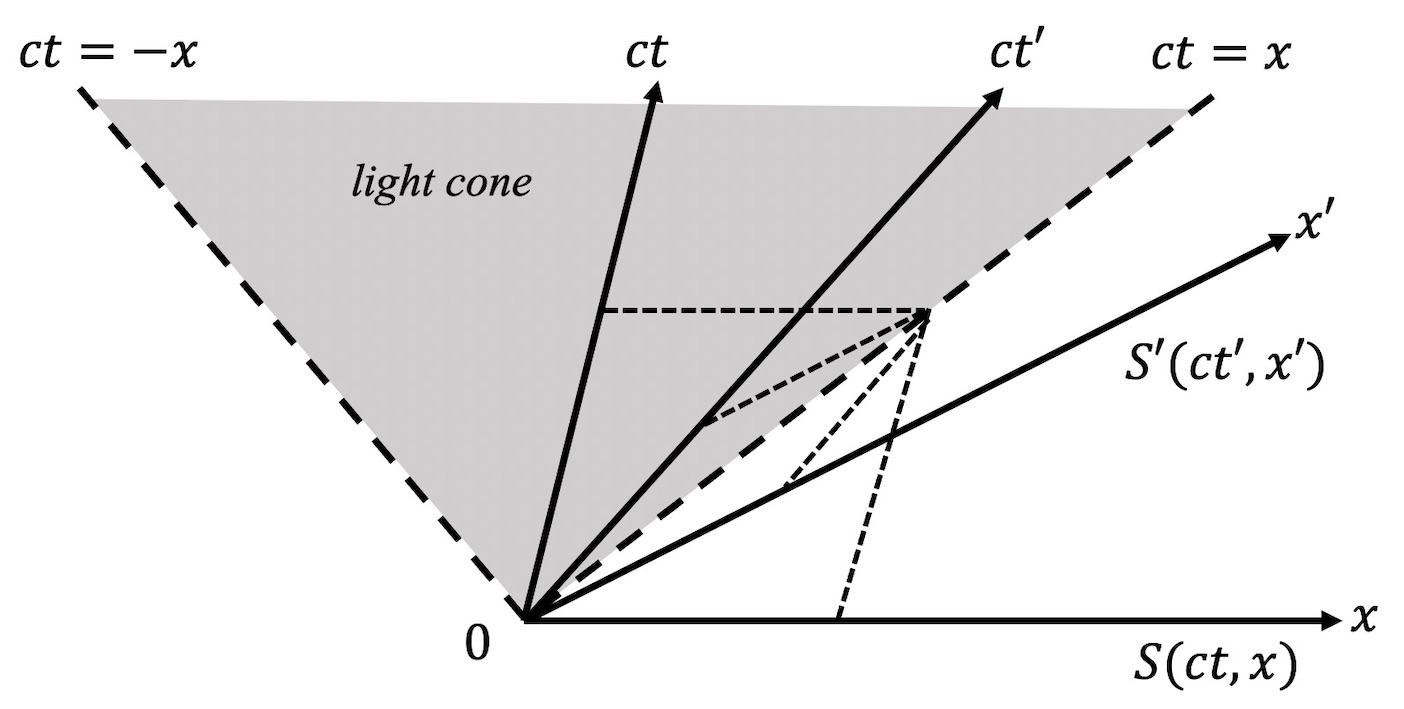}
\caption{Two reference systems, $S(ct,x)$ and $S'(ct',x')$, with the latter moving at coordinate velocity $V$ with respect to the former. The worldline of a photon (the dotted line) is then the bisector of the time axis $ct$ and space axis $x$, that is, the set of points such that their affine distances to the two axes are the same. Relative to the reference system $S(ct,x)$, the axis $ct'$ of the reference system $S'(ct',x')$ has equation $ct= {c\over V} x$. Hence, for the photon worldline to be the bisector of $ct'$ and $x'$ as well, the equation of the $x'$-axis must be $ct={V\over c}x$. It is important here to remember that this diagram, although inevitably plotted on a two-dimensional Euclidean support (the plane of the sheet of paper), it is an affine diagram, expressing a non-Euclidean metric space, where the coordinates along the different axes (which are not necessarily perpendicular) are obtained by projections parallel to the axes.
\label{figure2}}
\end{figure}

To further analyze the meaning of a movement also happening `in time', we first observe that the temporal dimension, when viewed as part of a spacetime, also possesses the dimension of a length, the time variable being multiplied by the speed $c$. In other words, time is here considered on a par with a spatial dimension. For example, if the time unit is taken to be one year, then the corresponding unit of length will be one \emph{light year}. More concretely, in addition to the movement introduced by Copernicus, i.e., a movement around the Sun, that the Earth performs in one year, in that same year Earth also moves a temporal distance which is approximately one light-year. In other words, although we have previously considered reference systems of the $\Sigma(t,x)$ kind, with $(t,x)$-variables, when we move to a discussion where the four-velocity is considered to be a relevant physical quantity, we must also consider the time axis as an axis having the dimension of a length, i.e., we must consider a dimensionally homogeneous reference system $S(ct,x)$, with $(ct,x)$-variables, in which a 
photon's worldline becomes the bisector between the temporal $ct$-axis and the space $x$-axis; see Figure~\ref{figure2}.

Our previous analysis also makes it clear that when describing the movement of an entity, we must always consider its proper time, and that if an entity $A'$ moves with respect to another entity $A$, say with coordinate velocity $V$, then it will move `in time' along a proper time direction that will be different from that along which entity $A$ moves `in time'. The existence of this multiplicity of different time directions, instead of a single global chronological time, can more easily be understood when adopting a geometric perspective. Imagine a spatial $(x,y)$-plane, equipped with the usual Euclidean metric. We know that we can draw an infinite number of lines passing from the origin of the chosen reference system, and that all these lines correspond to different spatial directions. In much the same way, a spatiotemporal $(ct,x)$-plane, equipped with Minkowski's metric, also has infinitely many lines passing through the origin and contained in the \emph{light cone}, which correspond to the different possible proper time directions, hence to the different possible worldlines; see Figure~\ref{figure2}.

\begin{figure}[!ht]
\centering
\includegraphics[scale =0.2]{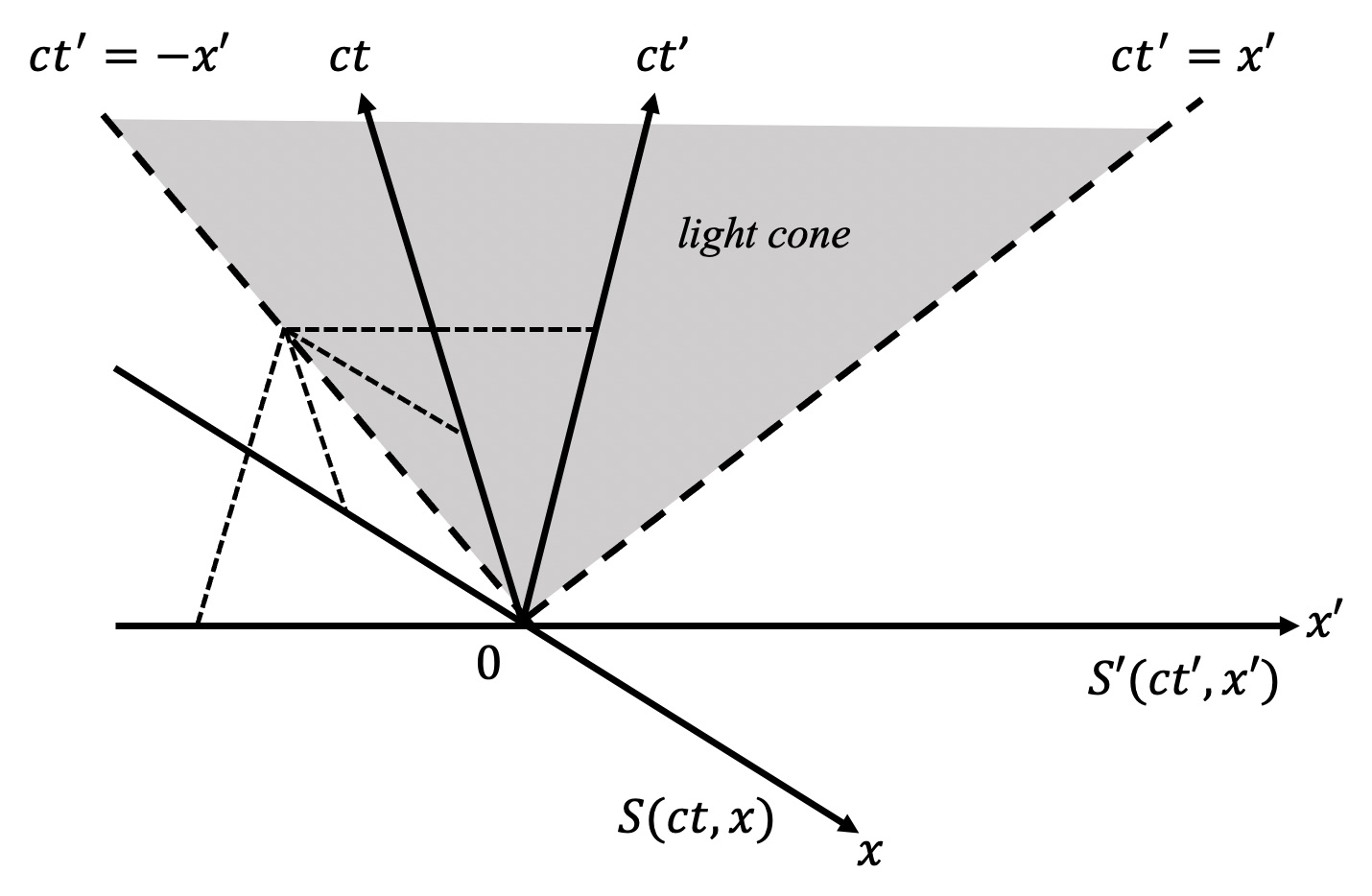}
\caption{Two reference systems, $S'(ct',x')$ and $S(ct,x)$, with the latter moving at coordinate velocity $-V$ with respect to the former. See also the caption of Figure~\ref{figure2}.
\label{figure3}}
\end{figure}

When we say that $A'$ moves relative to $A$ with coordinate velocity $V$, it means that we describe the movement of the centre of mass of $A'$ in the coordinate system of $A$, with $S(ct,x)$ the reference systems associated with $A$ and $S'(ct',x')$ the reference systems associated with $A'$; see Figure~\ref{figure2}. We can of course also consider the situation of the center of mass of the entity $A$ moving with coordinate velocity $-V$ relative to the coordinate system of $A'$; see Figure~\ref{figure3}. The first situation is a description in which the $x$-space points of $A$, which are the points of simultaneity with respect to $A$, are taken as the scene. The second situation is a description in which the $x'$-space points of $A'$, which are the points of simultaneity with respect to $A'$, are taken as the scene. These two scenes, precisely because of the non-existence of a common notion of simultaneity for $A$ and for $A'$, are very different from each other, and yet we still usually reason about them as if they were one and the same stage.

\begin{figure}[!ht]
\centering
\includegraphics[scale =0.2]{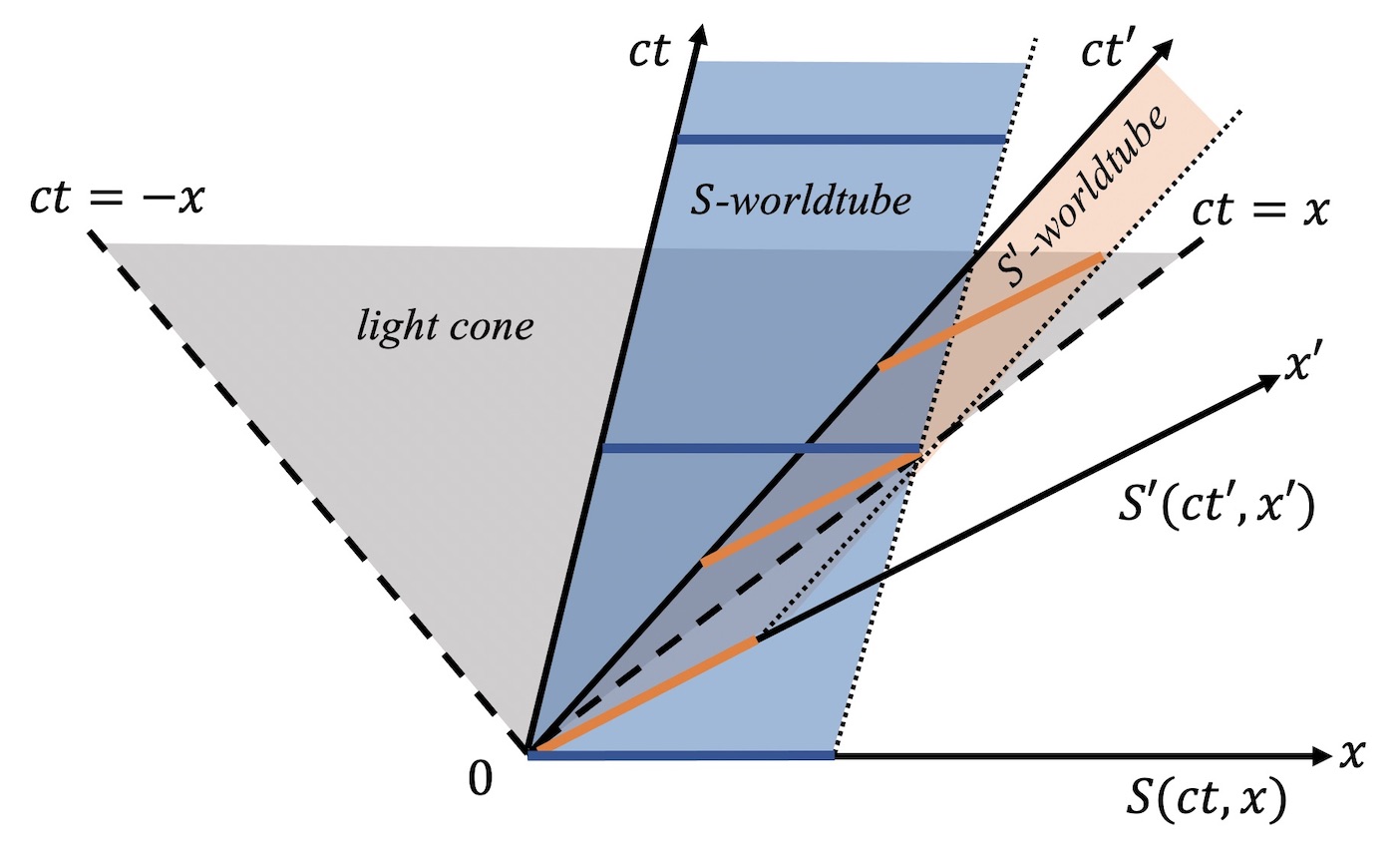}
\caption{Two identical rods (of different color but same proper length) are represented: the blue rod is at rest with respect to the reference system $S(ct,x)$, whereas the orange rod is at rest with respect to the reference system $S'(ct',x')$, which moves at (coordinate) velocity $V$ with respect to $S(ct,x)$; see also the caption of Figure~\ref{figure2}. The spatiotemporal orientations of the two rods are not the same, which gives rise to the relativistic effects of length contraction and time dilation. In the diagram, the worldtubes of the two rods are also represented. \label{cuboid1}}
\end{figure}

\begin{figure}[!ht]
\centering
\includegraphics[scale =0.2]{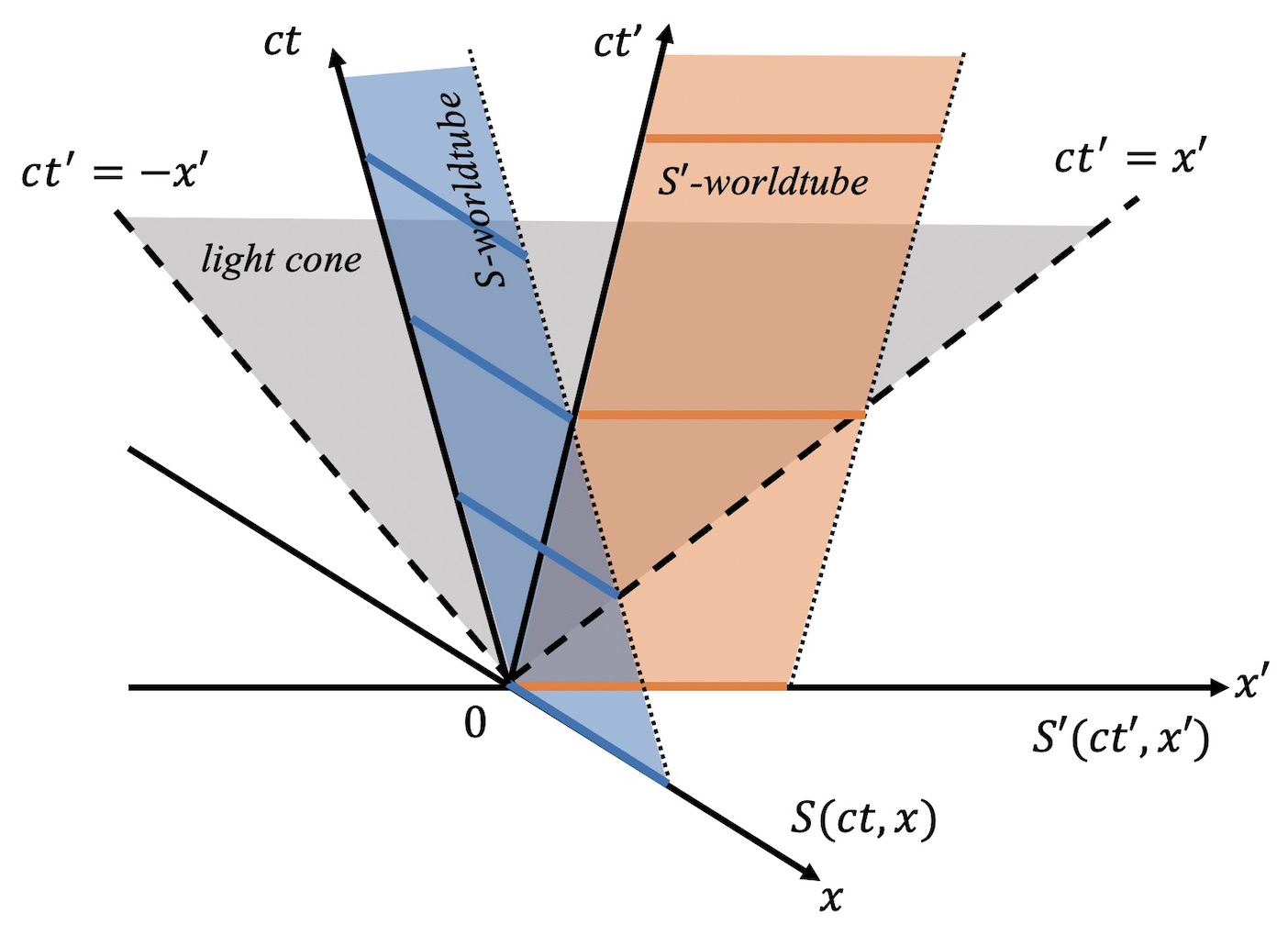}
\caption{Same situation as in Figure~\ref{cuboid1}, but with the reference system $S(ct,x)$ moving at coordinate velocity $-V$ with respect to the reference system $S'(ct',x')$; see also the caption of Figure~\ref{cuboid1}.
\label{cuboid2}}
\end{figure}

Now, as long as we are talking about point particles, or only considering the center of mass of entities, one can still maintain this illusion that there would be a single spatiotemporal scenery, but when macroscopic entities having a volume and a shape are involved, the exercise becomes much more difficult, for the appearance of an entity in one scene no longer corresponds to the same appearance in a different scene, because of how the relativistic effects act differently on the different elements forming the entity in question. To make this more concrete, consider a material entity having the shape of a cuboid. Since in our discussion we are only considering a single dimension of space, the cuboid will reduce to a one-dimensional rod. If we try to represent this rod similarly to what we did in Figures~\ref{figure2} and \ref{figure3} for a center of mass material entity, we obtain the diagrams of Figures~\ref{cuboid1} and \ref{cuboid2}. 

We see that a same rod occupies a very different set of spacetime points in the two experiencer's coordinate systems $S(ct,x)$ and $S'(ct',x')$. This is because the set of spacetime points occupied by the rod in one coordinate system are those that realize a condition of simultaneity there, which are not the same set of simultaneity points for the other coordinate system. Also, if one considers the worldlines of all spacetime points forming the rod, one obtains two different \emph{worldtubes}, in the two reference systems, corresponding to the two regions colored blue and orange in Figures~\ref{cuboid1} and \ref{cuboid2}. Hence, in the same way the hyperplanes of spatial simultaneity (which in our simplified situation are just lines) that are associated with a spatially extended entity are fundamentally different for experiencers moving relative to each other, the same is true for the corresponding worldtubes, which also present themselves in a very different way in the personal spaces of these experiencers. Hence, we can say that the typical block universe strategy which consists in considering the worldtubes as the elements of reality, i.e., as `what exists' in relativity, fails for the same reasons that simultaneity fails in relativity.

\section{Temporal energy}
\label{Temporal energy}

The perspective describing the movement of physical entities as happening not only in space, but also in time, allows to additionally explain the origin of Einstein's \emph{mass-energy equivalence} as a form of \emph{temporal energy}, and this is an extra argument for taking seriously the notion of four-velocity in special theory of relativity. To see this, let us start by recalling that the four-velocity retains, in the relativistic regime, the interpretation of being equal to the momentum of an entity per unit of its mass. Indeed, if we define the \emph{four-momentum} (which is a two-momentum, in our case) by $p_{\rm p}=m_0u_{\rm p}$, we have
\begin{equation}
p_{\rm p}= \left[\begin{matrix}p_{\rm p}^0\\p_{\rm p}^{\rm s}\\\end{matrix}\right]= m_0 \left[\begin{matrix}v_{\rm p}^0\\v_{\rm p}\\\end{matrix}\right]= m\left[\begin{matrix} c\\ v\\\end{matrix}\right]\quad\quad \|p_{\rm p}\|=m_0 c\quad\quad m=\gamma m_0
\label{4-momentum}
\end{equation}
where $m_0$ is the rest (proper) mass and $m$ is the (coordinate) relativistic mass.

We can then observe that the spatial component $p_{\rm p}^{\rm s}$ of the four-momentum is given by the usual relation of mass times velocity, which holds either for the proper space velocity, and then the rest mass has to be used, or for the coordinate velocity, and then the relativistic mass has to be used: $p_{\rm p}^{\rm s}=m_0 v_{\rm p} = m v$. 
On the other hand, the momentum time component, $p_{\rm p}^0 = mc=\frac{E}{c}$, corresponds to the energy of the entity in question divided by its spatiotemporal speed $c$ (which is also the coordinate speed of light), where $E$ is given by the famous Einsteinian formula $E=mc^2=m_0\gamma c^2$. In the limit $v_{\rm p}\to 0$, of an entity spatially at rest, we have $E = m_0c^2$, $p_{\rm p}^0=m_0c$ and $p_{\rm p}^{\rm s}=0$. This explains why all the energy contained in the entity's rest mass $m_0$ can be interpreted as the energy associated with its motion, at speed $c$, along it's time direction, i.e., as a form of \emph{kinetic temporal energy}.

What about these formulas for zero rest mass entities like photons?  If $m_0 = 0$, the temporal and spatial proper velocities are infinite and we have indeterminate expressions of the type ``zero times infinity." However, when $m_0 = 0$, we also get from (\ref{4-momentum}) that $\|p_{\rm p}\|=0$, which implies:
\begin{equation}
\|p_{\rm p}\|^2=(p_{\rm p}^0)^2-(p_{\rm p}^{\rm s})^2=\left(\frac{E}{c}\right)^2-(p_{\rm p}^{\rm s})^2=0
\label{momentum-light}
\end{equation}
from which we obtain $E=c\, p_{\rm p}^{\rm s}$, which expresses the correct relationship between the energy and spatial proper momentum of a photon. Of course, to get the more specific Planck-Einstein relations \citep{Planck1901}, $E=h\nu$, $p_{\rm p}=\frac{h}{\lambda}$ and $\nu\lambda=c$, which relate energy and momentum to the frequency $\nu$ and wavelength $\lambda$ associated with a photon, one needs to revert to quantum mechanics.

\section{Quantum and conceptuality} 
\label{Quantum and conceptuality}

Coming to quantum mechanics, our analysis shows that concepts that specifically came to prominence in quantum theory, already made their appearance in relativity theory, and one of these concepts is \emph{contextuality}. As we mentioned already, the fact that for each experiencer it is necessary to consider his/her personal spacetime, this is a form of contextuality, albeit of a specific relativistic type. As we noted, the operational analysis of `what exists' is inspired by our research in axiomatic quantum mechanics, where it is assumed that reality is non-deterministic and experiencers can make unpredictable choices \citep{Aerts1996,Aerts1999}. 

On the other hand, even more important  for the analysis we proposed in this article, is the quantum mechanical notion of \emph{non-locality}, which in our research group we view as an expression of \emph{non-spatiality}, a notion that one of us introduced as early as the late 1980s \citep{Aerts1990} and was discussed in a number of works \citep{Aerts1998,Aerts1999}. Of course, others have also subsequently realized its importance, like Ruth Kastner in her \emph{possibilist transactional interpretation} \citep{kastner2012,kastner2022,Aertssassolidebianchi2017}. In other words, for over forty years the idea has been proposed that quantum physics can only be understood if one accepts that our physical reality is essentially non-spatial, and more generally non-spatiotemporal \citep{Aerts1999,Sassolidebianchi2021}. 

It is important to underline, however, that the notion of non-spatiality was not introduced as a mere philosophical speculation, but as a necessary ingredient to really explain the behavior of quantum entities in key experiments. We have already mentioned in  Section~\ref{introduction} those about entanglement, but they were certainly not the only ones. Just to provide another important example, in the seventies of the last century a number of experiments were carried out with ultracold neutrons, using perfect silicon crystal interferometers, which allowed to test all sorts of quantum properties, like the $4\pi$-symmetry of spin-$1/2$ entities \citep{BonseRauch1979,Greenberger1983,Hasegawa2011,RauchWerner2015}. When these experiments are analyzed without preconceptions, it is clear that neutrons cannot be interpreted as spatial entities, be them localized or extended \citep{Aerts1999, Sassolidebianchi2017, Sassolidebianchi2021}.

More specifically, many of the mysterious quantum features, like \emph{superposition}, \emph{measurement}, \emph{entanglement}, \emph{complementarity} and \emph{indistinguishability}, can be considered to be an expression of the non-spatiotemporality of the quantum entities, i.e., of the fact that, generally speaking, the presence of a quantum entity within the spatiotemporal theater described by a given frame of reference is only of a \emph{potential} nature, hence, spacetime should be viewed not only as an emerging personal structure that can be associated to each macroscopic material entity, but also as a specific experimental context that can be concretely implemented by providing specific position measuring instruments. 

The question of the reality status of movement in time can also be better grasped if placed in the broader perspective of quantum mechanics. In our view, it is fair to say that in a context where spacetime has emerged, a motion in time is as real as a motion in space. Furthermore, when time and space are considered jointly, time appears to be more fundamental, the change we experience being more properly related to it, while space defines the scenery within which we can order change, not only our own but also that of other entities. 

The fourth temporal dimension, which in Minkowski's representation is also a spatial dimension, should not, however, be trivially identified with time as such, that is, with what is at the origin of change. And just as it is necessary to have a reference spatial domain to give meaning to the notion of spatial proper velocity, in the same way a reference time domain (associated with a clock) is necessary to give meaning to the notion of temporal proper velocity, which as we have seen is always greater than or equal to $c$, and similar to pre-Copernican times (see the discussion in the concluding section), the reason why we hadn't figured out this additional time velocity is because it isn't easy to experience it, just as it wasn't easy to experience the speed of Earth in space.

However, if there is a non-spatiotemporal domain underlying the spatiotemporal one, it remains an open question to know what \emph{change} means in that domain, while preserving the possibility of also explaining the kind of change that we all experience continuously and in a completely evident way. In this regard, we proposed the \emph{conceptuality interpretation} \citep{Aertsetal2018,AertsSassoli2022}, an interpretation of quantum mechanics, still in development, that gives a concrete expression to this non-spatiality and non-temporality. Of course, it is not possible in the limited space of this article to go into the merits of this interpretation, which we believe offers the missing ontology and metaphysics that can make quantum and relativity theory fully intelligible, by allowing to explain those key phenomena that in most interpretations remain unexplained. 

Its basic assumption is that quantum  entities, in their states that are subject to measures, are \emph{conceptual entities}, hence ontologically similar to human concepts (but not to be confused with the latter), in the sense that they are carrier of a substance, in the quantum jargon referred to as \emph{coherence}, which is similar to \emph{meaning}, and that measuring apparatuses behave similarly to cognitive entities that are sensitive to their meanings. Hence within this conceptuality interpretation the ontological nature of `object' for a quantum entity is put into question. 

To make the similarity with human language more concrete, consider the concept `horse'. Obviously, it is not an object, and only a very concrete instantiation of it, for example `one specific material horse standing in a meadow that can be petted', becomes an object. Similarly, a quantum entity would be conceptual in nature (although not a concept belonging to the human cognitive domain), thus capable of changing its state in becoming more objectual, i.e., more concrete, more localized in space, or more conceptual, i.e., more abstract, more de-spatialized, and this inevitable trade-off between concrete and abstract, between objectual and conceptual, would be nothing more than an expression of \emph{Heisenberg's uncertainty principle}.

We are also now in a position to answer the question with which we concluded Section~\ref{The future in the present}. We have placed a strong emphasis on the way in which the existence of free choice, for each experiencer, affects the nature of their personal block universes, specifically with the presence of bifurcations at each point of their worldlines, the structure of which expresses the reality of these free choices of the experiencers. Does this mean that in the absence of experiencers, that is, for example, in the period when human beings had not yet appeared on the Earth's surface, there were no bifurcations on the worldlines of physical entities? Certainly not. Perhaps it is not appropriate to speak of free choice here, but a physical entity composed of fermionic matter will be associated with the intrinsic and irreducible indeterminism that is manifested at the quantum level, when such a physical entity is used as a measuring instrument.

Just as we believe that relativistic effects exist whether or not there are experiencers to experience them, we also believe that the irreducible quantum indeterminism associated with the interactions of a physical entity exists, whether or not a physical entity is used as a measuring instrument. This is also understood in the assertion, in the conceptuality interpretation, that a piece of fermionic matter behaves as a cognitive entity and thus can play the role of an experiencer in relativity, giving rise to the presence of these bifurcations at all points along its worldline.

\section{Explaining Minkowski} 
\label{Explaining Minkowski}

Keeping the conceptuality interpretation of quantum mechanics in mind, in this section we indicate how the Minkoswki metric can be explained under the assumption that experiencers are cognitive entities making reasoning processes, and that the content of their reasonings, when going from a given \emph{hypothesis} to a given \emph{conclusion}, are the conceptual entities they interact with. Our additional assumption, which reflects the observation that the magnitude of the four-velocity is an invariant, is that they all reason at the same intrinsic speed, to be interpreted as the frequency with which they produce their elementary cognitive steps. 

\begin{figure}
\begin{center}
\includegraphics[scale =0.3]{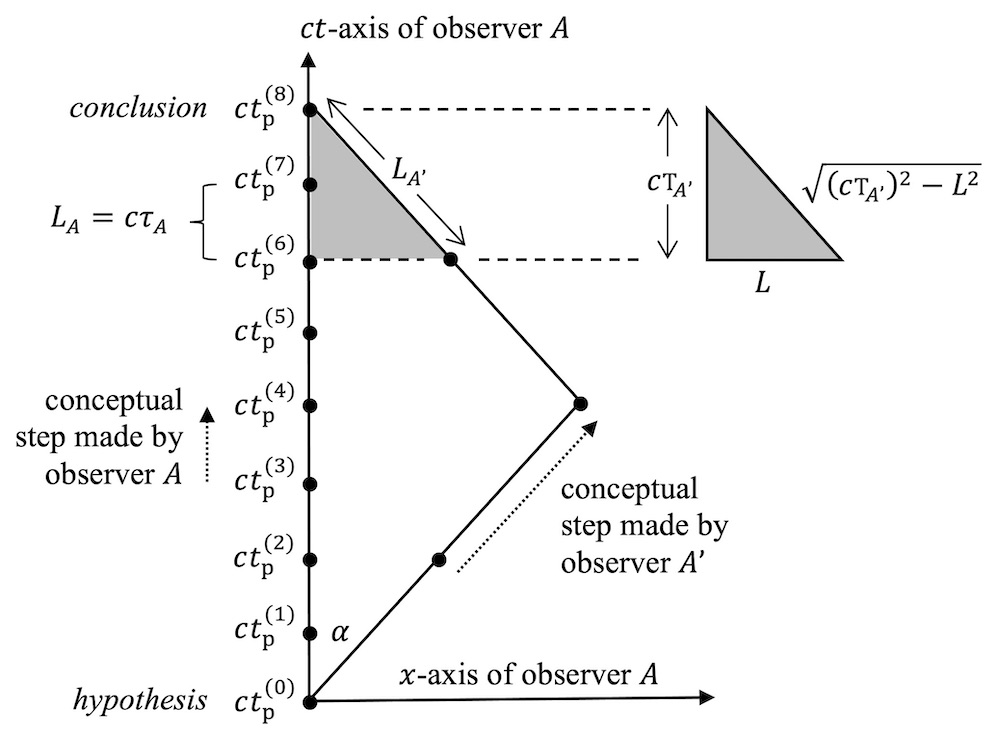}
\end{center}
\vspace{-15pt}
\caption{The two conceptual paths followed by experiencer $A$ and experiencer $A' $, in the personal spacetime $S(ct,x)$ of the former.} 
\label{figure6}
\end{figure}

More precisely, let us suppose that the reasoning process of experiencer $A$ happens in 8 conceptual steps, ranging from an initial \emph{hypothesis}, happening at personal time $t_{\rm p}^{(0)}$, to a certain \emph{conclusion}, happening at personal time $t_{\rm p}^{(8)}$. To order his/her process, experiencer $A$ will attribute a same length $L_A$ to all of its $8$ steps, which will be organized sequentially along an axis: his/her proper time axis; see Figure~\ref{figure6}. So, we are assuming that the speed with which the different steps are produced is always the same and would in fact correspond to $c$, according to our previous analysis. 

Imagine then that experiencer $A$ does not want to only describe his/her specific process, which produces a conclusion in $8$ steps, but is also interested in putting it in relation with other cognitive processes. Imagine a second process, produced by another cognitive entity, let us call him/her experiencer $A'$, who starts from the same hypothesis and reaches the same conclusion, but he/she does so in only $4$ steps, so, in a sense, experiencer $A'$ produces a more effective reasoning. 

When experiencer $A'$ focuses on his/her cognitive process, he/she will of course also introduce a personal time axis, to order his/her $4$ elementary steps, which will also take place at the speed of light. But how does experiencer $A$ also represent the reasoning of experiencer $A'$ in a coherent way, considering that in just $4$ steps he/she reaches the same conclusion, starting from the same hypothesis? Clearly, he/she cannot represent it directly on the axis that was built to order his/her $8$-step reasoning, because the steps of both experiencers, $A$ and $A'$, are elementary cognitive steps, produced at the exact same speed. To put it differently, they are steps of the same ``length,'' and the scale on the time axis of experiencer $A$ has been defined in such a way that, to go from the hypothesis to the conclusion, exactly $8$ elementary steps are needed, and not $4$.

Therefore, either experiencer $A$ renounces to relate his/her cognitive process with that of experiencer $A'$, or, if he/she wants to do so, he/she will have to introduce new Cartesian axes, with new parameters. And these would be axes of a spatial kind for experiencer $A$. So, in our situation, $A$ can simply build a new axis orthogonal to the first one, and describe the cognitive process of experiencer $A'$ as something that moves along the direction of such new axis, at a certain speed $v$, starting from the common point of the hypothesis, to then reverse the direction of travel to go back and meet again experiencer $A$, at the spatiotemporal point corresponding to the joint conclusion. 

Now, as can be seen on Figure~\ref{figure6}, there would be a problem if one tries to interpret this construction using the Euclidean geometry. Indeed, as we said, the two experiencers' elementary steps are perfectly equivalent, as they happen at the same speed, in the same abstract atemporal background. But according to the aforementioned construction, it would seem that the steps of experiencer $A'$ are longer. And this is where the Minkowskian, non-Pythagorean metric comes in, allowing the hypotenuse of a triangle to be shorter than one of its two catheti, so that the steps of experiencer $A'$ can become exactly of the same length as those of experiencer $A$.

More precisely, if $L$ is the component of the length of an elementary step $L_{A'}$, taken by experiencer $A'$ along the space axis of experiencer $A$, then according to the Minkowski metric we have (see Figure~\ref{figure6}): $L_{A'}^2 =(c\,\textsc{t}_{A'})^2 -L^2$, so that the requirement that $L_A=L_{A'}$, or equivalently $L_A^2=(c\,\tau_A)^2$, considering that $\tau_A=\textsc{t}_{A'}/\gamma$ and $c\,\tau_A= \sqrt{c^2-v^2}\,\textsc{t}_{A'}$, gives: $(c^2-v^2)\,\textsc{t}^2_{A'} =(c\,\textsc{t}_{A'})^2 -L^2$, that is, $L=v\textsc{t}_{A'}$. In other words, by adopting a pseudo-Euclidean metric, experiencer $A$ constructs a spacetime theater in which he/she can now consistently keep track not only of his/her cognitive processes, but also of those associated with experiencer $A'$. Of course, a single spatial axis will be sufficient when considering only two entities, but additional axes are needed if further entities are jointly considered \citep{Aerts2018}. 

It is of course a limitation of us humans to not be able to view the diagram of Figure~\ref{figure6} in the correct way, as we have evolved using bodies whose relative motions, when described in a spatiotemporal theater, are very slow relative to the speed of light. So, for us it was as if the construction of the time axis and of the spatial axes were perfectly independent, when instead they are intimately connected. In other words, since the relativistic effects were negligible for us, we have not incorporated them in our mental representation of the world.

\section{Concluding remarks} 
\label{Conclusion}

Summing up what we did, we have analyzed how an operational construction of reality allows us to overcome the limitations of the block view of the universe, by reintroducing change/creation into our description of the physical world, carefully distinguishing the observers, that we have more generally called experiencers,  from the spatiotemporal representations with which they can be associated, but of which they are not a part, in the same way that a reader walking between the lines of a book is not a part of its narrative. 

Thanks to this demarcation, it became possible to look at Langevin's twins' situation from a different perspective, extracting from it the information that each experiencer (and by extension each material entity) is associated with a personal spacetime that is a construction in which the future is also in the present, in the sense that it contains the worldtubes of all entities different from their own body. 

We also analyzed motion from a new perspective, emphasizing that it is the notion of proper velocity that allows to understand the counterintuitive aspects of the invariance of the speed of light. We emphasized that the proper velocity is a component of the so-called four-velocity, which if taken seriously reveals that motion manifests not only with respect to space, but also, and especially, with respect to time, with all physical entities moving with the same proper four-speed, which only incidentally is equal to the coordinated speed of light $c$. Therefore, the latter should be primarily understood as a velocity in time, since massive entities always move with that velocity along their personal time directions, when spatially at rest in their own frames of reference. And from that perspective, a spaceship should be primarily called a timeship!

In other words, motion would not be what we usually think it is, and we can say that we remain crypto-Newtonians in relativity textbooks, when we only connect $c$ to the coordinate speed of light, without saying that, in the first place, it describes a `speed in time', when entities are spatially at rest, and more precisely a `speed along the fourth dimension of a four-dimensional  
space', which only for historical reasons continues to be called a time dimension, as part of a spacetime.

In this regard, let us observe that the fourth dimension intervening in the Minkowski metric is really $ct$, and not $t$, a fact that we emphasized in Section~\ref{Multiplicity of times}, when we considered a dimensionally homogeneous spatial reference system $S(ct,x)$, with $(ct,x)$-variables, instead of a dimensionally inhomogeneous spatiotemporal system $\Sigma(t,x)$, with $(t,x)$-variables. This means that the block universe should really be understood as a space of four dimensions, and not as a spacetime. And clearly, nothing changes in a pure spatial domain, as things can change in `a space' only with respect to `a time'. This was true for the pre-relativistic Newtonian three-dimensional space and will remain true for the relativistic four-dimensional Minkowski space, which should really be called `a space' and not `a spacetime'. 

Now, if physical entities actually move in a four-dimensional Minkowskian space, which means that the latter is not just a mathematical artefact but a real structure emerging from a deeper non-spatial domain, as we emphasized in Section~\ref{Quantum and conceptuality}, we have to admit that the revolution initiated by Copernicus remains to be completed, since this extended four-dimensional spatial movement is still not properly taken into account today in our relativistic description of the motion of physical entities, while our analysis shows that this speed in time exists on a par with the speed in space revealed by the Copernican revolution.

To understand why this is the actual state of affairs, let us take a step back for a moment and return to the famous phrase anecdotally attributed to Galileo Galilei, after he was obligated to retract his claims that, following Copernicus, the motions appearing in the firmament were the consequence of Earth's motion: ``And yet it moves'' (``eppur si muove''). This phrase points to one of the difficulties in accepting the new Copernican view: that humans do not directly perceive the movement of the planet. If this was indeed the case, then of course Galileo's perspective would have not been controversial, but mere experimental evidence. What we consider as experimental evidence, however, also depends on our conceptualization of the world, so much so that ideas about what would be the effects of a moving Earth, say on the trajectory of projectiles, and the lack of observation of these effects, were considered at the time to be evidence of the falsity of the view of a planet moving around the Sun and spinning on its axis. 

What was underestimated was the fact that everything on the planet's surface was moving with it, including its inhabitants, and therefore the latter could not perceive with their senses, nor with elementary experiments, the planet's rotational motion, around itself and around the sun. Of course, following Galileo, we discovered other means to highlight the planet's movement, like observing the period of a Foucault pendulum, watching the planet rotate on itself directly from space, thanks to satellites, and for what concerns its revolution around the sun we can use phenomena such as stellar parallax, stellar aberration, and the Doppler effect. 

So, Copernicus, and later Galileo, revolutionized our view on movement, allowing us to become aware of the existence of movements that until then we were unaware of, and this not because they were hidden. As Edgar Allen Poe famously emphasized, the best place to hide something is often right out in the open. We humans were all openly moving together with the planet, but precisely because of that, we were not able to detect the planet's motion. Even though the planet motion is not a uniform one, our past blindness with respect to it was also in part related to the fact that, with good approximation, the planet's surface can be considered to be an inertial frame of reference, in the sense that in our day-to-day experiences the effects due to rotation, like the centrifugal and Coriolis fictitious forces, remain minuscule. And as Galileo himself observed, in his relativity principle \emph{per se}, inertial frames constitute non-trivial equivalent viewpoints on our physical world, where physical phenomena are perceived to be the same. 

Einstein's relativity is the next great revolution about motion, but similarly to Copernican revolution its acceptance does not appear to be easy, and it is the thesis we defend in this article that it has not been fully achieved, because what we physicists have not fully realized is that `Mikowski space' is as real as its little brother `Newton space', hence the material entities move much more and rather differently than the way Copernicus told us. The Polish astronomer changed our view by making us aware that Earth was actually also moving in space, but such movement would not be the end of the story, because Earth and all the other entities with a rest-mass different from zero would also move extremely swiftly along a fourth dimension. But again, since we all do so incessantly and all together, somehow similarly to what happened centuries ago with Earth's spatial movement, we do not perceive such additional movement and we can question its reality. Just as we can question the existence of a deeper non-spatial layer of reality where processes of change would actually occur.

This is not to say that there are no signs that would allow us to infer the existence of such motion along the fourth dimension, and as we pointed out in Section~\ref{Temporal energy}, one of these would be the famous mass-energy equivalence, which would be precisely the expression of a kinetic-type of energy along that additional spatial axis associated with the temporal direction. But apparently, these signs are currently being interpreted differently, so we can argue that Einstein’s relativity revolution has not yet been completed. Our hope is that this contribution of ours will help overcoming our residual pre-relativistic preconceptions and embracing the relativistic revolution more fully.

\end{document}